# Gamma-ray measurements of naturally occurring radioactive samples from Cyprus characteristic geological rocks

**Michalis Tzortzis and Haralabos Tsertos***

*Department of Physics, University of Cyprus, Nicosia, Cyprus*

**Stelios Christofides and George Christodoulides**

*Medical Physics Department, Nicosia General Hospital, Nicosia, Cyprus*

## Abstract

Using high-resolution γ–ray spectroscopy, the terrestrial gamma radiation in all the predominant types of geological rock formations appearing in Cyprus was measured. Soil samples were collected from each rock type, sealed in 1-*litre* plastic Marinelli beakers, and measured in the laboratory for 24 *hours* each. From the measured γ–ray spectra, activity concentrations were determined for $^{232}$Th (range from 1.3 to 52.8 $Bq\ kg^{-1}$), $^{238}$U (from 0.9 to 90.3 $Bq\ kg^{-1}$) and $^{40}$K (from 13 to 894 $Bq\ kg^{-1}$). Elemental concentrations mean values of (2.8 ± 0.7) *ppm*, (1.3 ± 0.3) *ppm* and (0.6 ± 0.1) % were extracted, for thorium, uranium and potassium, respectively. Absorbed dose rates in air outdoors were calculated to be in the range of 0.1–50 $nGy\ h^{-1}$, depending on the geological features, with an overall mean value of (14.7 ± 7.3) $nGy\ h^{-1}$. The corresponding effective dose rates per person outdoors were estimated to be between 0.1 and 61.4 $\mu Sv\ y^{-1}$, assuming a 20% occupancy factor.

* *Corresponding author. E-mail address: tsertos@ucy.ac.cy , Fax: +357-22339060*
*Department of Physics, University of Cyprus, P.O.Box 20537, 1678 Nicosia, Cyprus.*

## 1. Introduction

Natural radioactivity is composed of the cosmogenic and primordial radionuclides. Cosmogenic radionuclides, such as $^{3}$H, $^{7}$Be, $^{14}$C, and $^{22}$Na, are produced by the interaction of cosmic-ray particles (mainly high-energetic protons) in the earth's atmosphere. Primordial radionuclides (also called terrestrial background radiation) are formed by the process of nucleosynthesis in stars. Only those radionuclides with half-lives comparable to the age of the earth, and their decay products, can still be found today on earth, e.g. $^{40}$K, and the radionuclides from the $^{238}$U and $^{232}$Th series. Gamma radiation from these radionuclides represents the main external source of irradiation of the human body.

The external radiation exposure arises mainly from cosmic rays and from terrestrial radionuclides occurring at trace levels in all soils. While absorbed dose rate in air from cosmic radiation outdoors at sea level is about 30 nGy $h^{-1}$ for the southern hemisphere (UNSCEAR Report, 2000), the specific levels due to terrestrial background radiation are related to the types of rock from which the soils originate. Therefore, the natural environmental radiation mainly depends on geological and geographical conditions (Florou and Kritidis, 1992). Higher radiation levels are associated with igneous rocks, such as granite, and lower levels with sedimentary rocks. There are exceptions, however, as some shales and phosphate rocks have relatively high content of radionuclides (UNSCEAR Report, 1993).

The island of Cyprus has attracted a lot of geological interest because it contains a mafic intrusive complex that is thought to represent an excellent example of an ophiolite slice (Robinson and Malpas, 1998). The Cyprus ophiolite complex, known as Troodos Massif, consists of pillow basalts overlying a sheeted dyke complex, and then an intrusive complex of gabbro that grades downward into olivine gabbro, then into an intramafic body of harzburgite and dunite. Some of the harzburgites are serpentinized. The upper parts of the gabbro and the lower basalts are cut by many closely spaced diabase dykes that form conspicuous sheeted masses. They are overlaid with fine-grained ferruginous, siliceous and sulfide bearing sediments called umpres.

Investigations on terrestrial natural radiation have received particular attention worldwide and led to extensive surveys in many counties (see UNSCEAR 2000 Report and further references cited therein). They mainly serve as baseline data of natural radioactivity such that man made possible contaminations can be detected and quantitatively determined. They can further be used to assess public dose rates and to perform epidemiological studies. The results obtained in each country can be exploited to enrich the world's data bank, which is highly needed for evaluating worldwide average values of radiometric and dosimetric quantities (Al-Jundi, 2002).

There are no systematic data on this subject available for Cyprus. Only a few studies on natural radioactivity and on indoor radon concentration measurements are reported by two of the authors (Christofides, 1994; Christofides and Christodoulides, 1993). A pilot project was therefore initiated aiming at systematically measuring the terrestrial gamma radiation in Cyprus and at determining its contribution to the annual effective dose equivalent to the population. Other part of the project includes measurement of radon concentration in houses and public buildings (Anastasiou et al., 2002) as well as the detection of $\alpha$–emitting radioisotopes by utilizing radioanalytic techniques and high-resolution $\alpha$-spectroscopy. In this paper, the first results from natural gamma radiation measurements in all the types of geological rock formations appearing in Cyprus (Figure 1) are presented. The experiments have been carried out in the Nuclear Physics Laboratory of the Department of Physics, University of Cyprus, using a high-resolution γ–ray spectroscopic system.

## 2. Experimental method

### 2.1 Gamma-ray detection system

A stand-alone high-resolution spectroscopic system is used for the measurement of the energy spectrum of the emitted gamma rays in the energy range between 50 *keV* and 3000 *keV*. The system consists of a high-purity germanium (HPGe) detector (coaxial cylinder of 55 *mm* in diameter and 73 *mm* in length) with an efficiency of 30%, relative to a 3″×3″ NaI(Tl) scintillator. The detector is mounted on a cryostat which is dipped into a 30-*litre* dewar filled with liquid nitrogen. Experimental arrangement for spectra collection includes the high-voltage detector bias supply, equipped with a remote shutdown feature, and signal-processing electronics. The latter include a spectroscopy main amplifier which incorporates an efficient pile-up rejector, and a Multi-Channel Buffer (MCB) which is a PC-based plug-in PCI card consisting of an 8k ADC. An advanced Multi-Channel Analyzer (MCA) emulation software (MAESTRO-32) allows data acquisition, storage, display and online analysis of the acquired γ–spectra.

The detector is surrounded by a graded–Z cylindrical shield consisting of lead, iron, and aluminum with thickness of 5 *cm*, 1 *cm*, and 1 *cm*, respectively, which provides an efficient suppression of background gamma radiation present at the laboratory site. The energy-dependent detection efficiency has been determined using a calibrated $^{152}Eu$ gamma reference source (standard Marinelli beaker with 85 mm bore diameter), which has an active volume of 1000 *ml* and an average density of 1 $g\ cm^{-3}$. The energy resolution (FWHM) achieved is 1.8 *keV* at the 1.33 *MeV* reference transition of $^{60}$Co.

## 2.2 Sample collection and counting

According to the geological features of Cyprus shown in Fig.1, samples have been carefully collected from each type of the geological formations. Samples from different places of the same rock formation have been mixed together. A total of 28 samples have been collected (Table 1) which represent the predominant rock formations of Cyprus*. Measured samples were sieved through 0.2-*mm* mesh, sealed in standard 1000 *ml* Marinelli beakers, dry-weighed and stored for at least four weeks before counting in order to allow the in-growth of uranium and thorium decay products and achievement of equilibrium for $^{238}$U and $^{232}$Th with their respective progeny (Hamby and Tynybekov, 2000; Al-Jundi, 2002). However, significant disequilibrium is uncommon in rocks older than $10^6$ years, and the $^{232}$Th series may be considered in equilibrium in most geological environments (Chiozzi et al., 2002). Each sample was put into the shielded HPGe detector and measured for 24 hours (a typical spectrum is shown in Fig. 2). Prior to the samples measurement, the environmental gamma background at the laboratory site has been determined with an empty Marinelli beaker under identical measurement conditions. It has been later subtracted from the measured γ–ray spectra of each sample.

Each measured γ–ray spectrum has been analyzed offline by a dedicated software program (GammaVision-32), which performs a simultaneous fit to all the significant photopeaks appearing in the spectrum. Menu-driven reports are available for summaries including

* Samples from the geological formations consisting the Pentadaktylos mountains in the northern part of Cyprus could not be collected, because of the division of the island by force since 1974 (see Fig. 1).

centroid channel, energy, net area counts, background counts, intensity and width of identified and unidentified peaks in the spectrum, as well as peak and average activity in *Bq kg*$^{-1}$ for each detected radionuclide. The results of reported activity concentrations obtained for each of the measured samples and average activity for samples of ophiolitic and sedimentary origin are summarized in table 1.

## 2.3 Calculation of elemental concentrations

Following the spectrum analysis, count rates for each detected photopeak and activity per mass unit (specific activity or radiological concentration) for each of the detected nuclides are calculated. The specific activity (in *Bq kg*$^{-1}$), $A_{Ei}$, of a nuclide *i* and for a peak at energy *E*, is given by:

$$A_{Ei} = \frac{N_{Ei}}{\varepsilon_E \times t \times \gamma_d \times M_s} \tag{1}$$

where $N_{Ei}$ is the Net Peak Area of a peak at energy *E*, $\varepsilon_E$ is the detection efficiency at energy *E*, *t* is the counting livetime, $\gamma_d$ is the number of gammas per disintegration of this nuclide for a transition at energy *E*, and $M_s$ is the mass in *kg* of the measured sample. If there is more than one peak in the energy analysis range for a nuclide, then an attempt to average the peak activities is made. The result is then the weighted average nuclide activity.

Based on the measured γ-ray photopeaks, emitted by specific radionuclides in the $^{232}$Th and $^{238}$U decay series and in $^{40}$K, their radiological concentrations in samples collected were determined. Calculations relied on establishment of secular equilibrium in the samples, due to the much smaller lifetime of daughter radionuclides in the decay series of $^{232}$Th and $^{238}$U. More specifically, the $^{232}$Th concentration was determined from the average concentrations of $^{212}Pb$ and $^{228}Ac$ in the samples, and that of $^{238}U$ was determined from the average concentrations of the $^{214}Pb$ and $^{214}Bi$ decay products. Thus, an accurate measurement of $^{232}Th$ and $^{238}U$ radiological concentrations was made, whereas a true measurement of $^{40}K$ concentration was achieved.

Radiological concentrations of $^{232}Th$, $^{238}U$ and $^{40}K$ were then converted into total elemental concentrations of thorium, uranium and potassium, respectively, according to the following expression:

$$F_E = \frac{M_E \cdot C}{\lambda_E \cdot N_A \cdot f_{A,E}} \cdot \frac{1}{n}\sum_{i=1}^{n} A_i \qquad (2)$$

where $F_E$ is the fraction of element $E$ in the sample, $M_E$ is the atomic mass *(kg mol$^{-1}$)*, $\lambda_E$ is the decay constant *($s^{-1}$)* of the parent radioisotope, $N_A$ is Avogadro's number (6.023×10$^{23}$ *atoms mol$^{-1}$)*, $f_{A,E}$ is the fractional atomic abundance of $^{232}Th$, $^{238}U$ or $^{40}K$ in nature, $C$ is a constant (with a value of 100 or 1,000,000) that converts the ratio of the element's mass to soil mass into a percentage or *ppm*, and $A_i$ is the radiological concentration of $^{40}K$ (n=1) or that of selected daughter radionuclides in the decay series of $^{232}Th$ and $^{238}U$ (n=2). Total elemental concentrations are reported in units of parts per million (*ppm*) for thorium and uranium, and of percent (%) for potassium.

## 2.4 Derivation of the gamma dose rates in air outdoors

If natural radioactive nuclides are uniformly distributed in the ground, dose rates at 1 *m* above the ground surface are calculated by the following formula (Kohshi et al., 2001):

**Dose rate (*nGy h$^{-1}$*) = Concentration of nuclide (*Bq kg$^{-1}$*)** (3)

**× Conversion factor (*nGy h$^{-1}$ per Bq kg$^{-1}$*)**

The methodology used for the derivation of the gamma dose rates was introduced by Beck and De Planque (1968) and Beck et al. (1972). They used the polynomial expansion matrix equation method for solving the soil/air transport problem in order to calculate the exposure rates 1 m above ground level for distributed sources of gamma emitters in soil. Kocher and Sjoreen (1985) using the point-kernel integration method have also calculated conversion factors (absorbed dose rate in air per unit activity per unit of soil mass, $nGy\ h^{-1}$ *per* $Bq\ kg^{-1}$). In the 1990’s Monte Carlo techniques have been exploited almost exclusively in order to calculate absorbed dose in air. More complete studies for the calculation of conversion factors were performed by Chen (1991), K. Saito and P. Jacob (1995) and recently by Clouvas et al. (2000). In these studies different Monte Carlo codes were used, such as the MCNP code of Los Alamos (1986), the GEANT code of CERN (1993) and a Monte Carlo code (called MC) developed in the Nuclear Technology Laboratory of the Aristotle University of Thessaloniki, Greece.

The absorbed gamma dose rate in air is mainly determined from $^{212}Pb$ (photon energy 238 *keV),* $^{208}Tl$ (583 *keV),* $^{212}Bi$ (727 *keV)* and $^{228}Ac$ (911 *keV)* for the $^{232}Th$ series, and from $^{214}Pb$ (352 *keV)* and $^{214}Bi$ (609 *keV)* for the $^{238}U$ series. Due to the small fractional yield and relatively low energy of the emitted photons, radionuclides such as $^{226}Ra$ and $^{224}Ra$ can be neglected (Clouvas et al., 2001). In addition, the well-known interference between the gamma line of $^{226}Ra$ (186.2 *keV)* and $^{235}U$ (185.7 *keV)* is inevitable, especially in the presence of a relatively high uranium concentration (Canet, 1990) and, therefore, the above-mentioned line was not used for the determination of $^{238}U$. No other than naturally occurring radionuclides were detected in the measured samples, with the exception of a small contamination of $^{137}$Cs due to the Chernobyl nuclear accident. Its contribution to the total dose is insignificant and has been neglected.

The dose contribution of one photon energy to the total dose rate due to all photon energies emitted from a specific radionuclide can be easily deduced from the above Monte Carlo simulations. In table 2, the dose rate conversion factors (DRCF) for different radionuclides are summarized, as deduced by Beck et al., Saito and Jacob and Clouvas et al. They have to be multiplied by the measured radiological concentration in order to deduce the dose rate due to the entire series. It should be pointed out here that, using this calculation, the dose rate for $^{232}$Th and $^{238}$U series is the average for the different dose rates deduced by the different gamma peaks belonging to each series and not the sum. For $^{40}$K the dose rate conversion factor is simply the value that should be multiplied by the measured concentration at the energy of 1460 *keV*, in order to deduce the dose rate due to $^{40}$K.

The comparison between the results obtained by the three Monte Carlo codes and the results obtained previously by others indicate a relative agreement (less than 15% of difference) for photon energies above 1500 *keV*. However, the agreement is not so good (difference of 20-30%) for the low-energy photons (e.g. lower than 200 *keV)*. The differences between the results of the three Monte Carlo codes and the results obtained by other researchers, presented in table 2, are 15% for $^{232}$Th series, 18% for $^{238}$U series and 10% for $^{40}$K. Clouvas et al. point out that the most accurate results are obtained by the MCNP code, probably due to the use of the point detector by the MCNP. In the present work, the considered dose rate conversion factor for $^{232}$Th and $^{238}$U series, and for $^{40}$K, used in all dose rate calculations, is just the average of the total value for each series given by the three Monte Carlo codes. Calculations were done for every identified element and the average value for each series was considered.

Finally, in order to make a rough estimate for the annual effective dose outdoors, one has to take into account the conversion coefficient from absorbed dose in air to effective dose and the outdoor occupancy factor. In the UNSCEAR recent reports (1993, 2000), the Committee used $0.7\ Sv\ Gy^{-1}$ for the conversion coefficient from absorbed dose in air to effective dose received by adults, and 0.2 for the outdoor occupancy factor. Effective dose rate outdoors in units of *μSv per year* is calculated by the following formula:

$$\textbf{Effective dose rate } (\mu \text{Sv y}^{-1}) = \textbf{Dose rate } (\text{nGy h}^{-1}) \times 24\text{ h} \times 365.25\text{ d} \times 0.2 \text{ (occupancy factor)} \times 0.7 \text{ Sv Gy}^{-1} \text{ (conversion coefficient)} \times 10^{-3} \quad (4)$$

In table 3 the results obtained for the dose outdoors and the corresponding effective dose assessment for the $^{232}$Th and $^{238}$U series, and $^{40}$K for sample 23 (Montmorilonite – Bentonitic clay) are displayed.

## 3. Results and discussion

As shown in table 1, activity concentrations of $^{232}$Th ranged from 1.3 to 52.8 $Bq\ kg^{-1}$, of $^{238}$U from 0.9 to 90.3 $Bq\ kg^{-1}$ and of $^{40}$K from 13 to 894 $Bq\ kg^{-1}$. Of all the 28 samples measured in this study, Melange and Montmorilonite appear to have the highest concentrations of $^{232}$Th, whereas Raised Marine Terrace Deposits in Klavdhia village and Celestite exhibit the highest concentration of $^{238}$U. Troodos Upper Pillow Lavas (Olivine and Picrite Basalts) appear to have much higher concentration of $^{40}$K, when compared with the concentrations of all the other samples, reaching levels up to 894.2 $Bq\ kg^{-1}$. In general, activity concentration of $^{232}Th$ and $^{238}U$ is rather low in all the samples measured; this is particularly true in ophiolites, which show the lowest concentration of those radionuclides. Average concentration of $^{232}Th$ and $^{238}U$ in samples of ophiolitic origin is just (2.1 ± 0.2) and (6.5 ± 4.4) $Bq\ kg^{-1}$, respectively, whereas in sediments the corresponding values increase to (16.7 ± 3.7) and (21.0 ± 5.5) $Bq\ kg^{-1}$, respectively. In addition, average concentration of $^{40}K$ is (147.3 ± 79.3) $Bq\ kg^{-1}$ for ophiolites and (197.5 ± 35.2) $Bq\ kg^{-1}$ for samples of sedimentary origin. Calculated median values in all samples of both ophiolitic and sedimentary origin are: 11, 15 and 178 $Bq\ kg^{-1}$, for $^{232}Th$, $^{238}U$ and $^{40}K$ concentrations, respectively, while revised median values worldwide (UNSCEAR Report, 2000) are: 30, 35 and 400 $Bq\ kg^{-1}$, respectively. This reveals that the mean concentration levels measured in Cyprus from naturally occurring radioisotopes are at least by a factor of two lower than the corresponding values obtained worldwide.

The calculated elemental total concentrations of thorium, uranium and potassium are plotted in Fig. 3 for our twenty-eight samples collected from all over the island. Measured arithmetic mean concentrations over all samples are: (2.8 ± 0.7) *ppm*, (1.3 ± 0.3) *ppm* and (0.6 ± 0.1)% for thorium, uranium and potassium, respectively. For comparison, Myrick et al. (1983) have determined arithmetic mean concentrations and standard deviations of thorium and uranium in surface soils in more than 300 samples obtained from locations around the United States. The corresponding values are: (8.9 ± 4.2) *ppm* and (3.0 ± 2.5) *ppm*, respectively. Also, Chang et al. (1974) reported the concentrations of thorium, uranium and potassium in earthen building materials of Taiwan to range from 14 to 16 *ppm*, 1.2 to 4.3 *ppm* and 0.15 to 12.8%, respectively. Potassium concentrations in a wide variety of rock types are estimated to range from approximately 0.1 to 3.5% (Kohman and Saito, 1954). Very recently, Chiozzi et al. (2002) have performed a series of investigations on different rock types at the Alps-Apennines transition, and found elemental concentrations being in the range: 0.3–16.7 *ppm*, 0.3–5.6 *ppm*, and 0.14–5.14 %, for thorium, uranium and potassium, respectively. Our extracted values (see Fig. 3) nicely fit to these results and, in general, they fall within the range of most reported values from other worldwide areas.

In order to establish a relation between the soil nature and the natural gamma dose rate in air, spectra were collected from several samples representing the predominant types of rocks in Cyprus. Characteristic rock types can be separated into two main categories: those that belong to the ophiolitic complex of Troodos (plutonic rocks, intrusive rocks – dykes and extrusive rocks – lavas) and those of sedimentary origin. Ophiolitic origin rocks could be characterized as undersaturated because they were formed by magma crystallization, which begins on high melting point ferromagnesian silicates and calcic plagioclases poor in silica. Ophiolitic rock types are mostly iron and magnesium saturated or oversaturated and this allows the characterization of these types as basic and superbasic. None of these rock types belongs to the category of silica-oversaturated rocks, which usually exhibit a predominant contribution of $^{238}U$ series elements. For rock types of sedimentary origin, a predominant contribution of $^{238}U$ series elements was found, followed by $^{40}K$ and $^{232}Th$ series elements. Both natural chemical rocks like limestones etc. and terrigenous sediments like sandstones etc. were considered, although the chemical and mechanical weathering of original rocks makes difficult their association.

After the data analysis obtained for rocks of ophiolitic origin, a predominant contribution of $^{40}K$ was found, mainly at the samples stemming from Troodos Upper and Lower Pillow Lavas. Other samples such as Mathiatis Pyrite Mine Tippings show a strong contribution of

$^{238}U$ series elements, but none of the samples shows a significant contribution due to the daughter elements of the $^{232}Th$ series. Rocks of ophiolitic origin present a major contribution to dose from $^{40}K$, which is significantly higher than those due to $^{232}Th$ and $^{238}U$ series. The dose rates in air outdoors were calculated from concentrations of nuclides of $^{232}Th$ and $^{238}U$ series, and of $^{40}K$ using equation (3). In the rocks of ophiolitic origin, the extracted values ranged from 0.1 to 0.6 $nGy\ h^{-1}$ with a mean of (0.2 ± 0.1) $nGy\ h^{-1}$, from 0.1 to 19.1 $nGy\ h^{-1}$ with a mean of (2.1 ± 1.7) $nGy\ h^{-1}$, and from 0.1 to 33.8 $nGy\ h^{-1}$ with a mean of (5.1 ± 3.1) $nGy\ h^{-1}$, respectively.

For sedimentary origin rocks, the contribution to dose due to $^{238}U$ series elements is the most important, followed by those from $^{40}K$ and $^{232}Th$ series elements. The corresponding dose rates calculated from concentrations of nuclides of the $^{232}Th$ and the $^{238}U$ series, and of $^{40}K$, in the rocks of sedimentary origin, ranged from 0.1 to 18.3 $nGy\ h^{-1}$ with a mean of (5.1 ± 1.4) $nGy\ h^{-1}$, from 1.2 to 34.3 $nGy\ h^{-1}$ with a mean of (7.6 ± 2.1) $nGy\ h^{-1}$, and from 0.1 to 16.2 $nGy\ h^{-1}$ with a mean of (6.5 ± 1.4) $nGy\ h^{-1}$, respectively. The total dose rates in air outdoors calculated for rocks of ophiolitic and sedimentary origin ranged from 0.1 to 34.8 $nGy\ h^{-1}$ with a mean of (7.4 ± 4.1) $nGy\ h^{-1}$ and from 0.9 to 48.7 $nGy\ h^{-1}$ with a mean of (19.7 ± 3.8) $nGy\ h^{-1}$, respectively. Figure 4 illustrates the measured relative contribution to the total dose rate outdoors for the two main rock type categories indicated. For rocks of ophiolitic origin, the relative contribution to dose due to $^{40}K$ is 68%, followed by a lower contribution due to $^{238}U$ and $^{232}Th$ series elements (29% and 3%, respectively). For sedimentary origin rocks, dose contribution levels due to $^{40}K$, $^{232}Th$ and $^{238}U$ series are comparable (34%, 27% and 39%, respectively).

The total mean absorbed dose rate in air outdoors from terrestrial radiation calculated for both rock types of ophiolitic and of sedimentary origin has a value of $(14.9 \pm 7.3)$ $nGy\ h^{-1}$. In the "Survey of Natural Radiation Exposures", the UNSCEAR refers especially for Cyprus an average absorbed dose in air outdoors of 18 $nGy\ h^{-1}$ with a range from 9 to 52 $nGy\ h^{-1}$ (Christofides, 1994). According to the recent UNSCEAR reports (1993, 2000), the corresponding worldwide average values range from 18 to 93 $nGy\ h^{-1}$ and a typical range variability for measured absorbed dose rates in air outdoors is from 10 to 200 $nGy\ h^{-1}$. The population-weighted values give an average absorbed dose rate in air outdoors from terrestrial gamma radiation of 60 $nGy\ h^{-1}$. It should be noted here that in the UNSCEAR reports, dose rate conversion factors are taken from K. Saito and P. Jacob (1995), which are 10–20 % higher than the more accurate values obtained recently by various Monte Carlo techniques (see table 2) and adopted in this work. The effective dose rates outdoors estimated according to equation (4) for rocks of ophiolitic and sedimentary origin ranged from 0.1 to 42.7 $\mu Sv\ y^{-1}$ with a mean of $(10.1 \pm 5.0)$ $\mu Sv\ y^{-1}$ and from 1.1 to 61.4 $\mu Sv\ y^{-1}$ with a mean of $(27.5 \pm 5.0)$ $\mu Sv\ y^{-1}$, respectively.

Finally, an attempt to find a relation between terrestrial radiation doses in air outdoors and the nature of the rock, studying dose contributions of each natural series, $^{232}Th$ and $^{238}U$, and $^{40}K$ that occur with different concentrations on rock composition. For rocks of ophiolitic origin the highest dose rate is 34.8 $nGy\ h^{-1}$ (42.7 $\mu Sv\ y^{-1}$) and was registered for Troodos Upper Pillow Lavas, where the relative weight to dose is 96.9% for $^{40}K$, 1.7% for $^{238}U$ series and 1.3% for $^{232}Th$ series. The lowest dose rate that registered was 0.1 $nGy\ h^{-1}$ for Harzburgite and Wehrlite. For rocks of sedimentary origin, the highest dose rate was 50 $nGy\ h^{-1}$ (61.4 $\mu Sv\ y^{-1}$) for Raise Marine Terrace Deposits in Klavdhia village, where the relative weight to the dose is 70.6% for $^{238}$U series, 16.4% for $^{40}$K and 13% for $^{232}$Th series. The lowest dose rate registered was 0.9 $nGy\ h^{-1}$ for Gypsum, and the relative weight is 90% for $^{238}$U series and 10% for $^{232}$Th series.

## 4. Conclusions

High-resolution γ–ray spectroscopy is a powerful experimental tool in studying natural radioactivity and determining elemental concentrations and dose rates in various rock types. For the predominant rock types of Cyprus that were investigated, samples originated from Troodos ophiolitic complex appear generally to have lower radionuclide concentrations, as compared to those of sedimentary origin. The latter, in combination with phosphates, carbonates and silicates, also present in sedimentary rocks, reveal a more significant concentration, especially in uranium and thorium series elements. However, activity and elemental concentrations of thorium, uranium and potassium in the studied soil samples are found to be normal. The extracted values are distinctly lower than the corresponding ones obtained from other countries worldwide and, in general, they all fall within the range given in the UNSCEAR Report (2000).

The mean absorbed dose rate in air outdoors amounts to (14.7 ± 7.3) $nGy\ h^{-1}$, which is by far below the corresponding population-weighted (world-averaged) value of 60 $nGy\ h^{-1}$. This implies that inhabitants of the island are subjected to a radiation exposure, which is significantly lower than the corresponding exposure levels reported in other areas worldwide. More systematic studies of the island's geological formations are in progress with the objective to create a digital radiological map of Cyprus, that would express the exposure levels due to terrestrial gamma radiation, which mainly depends on the geological features of Troodos ophiolite and the other sedimentary rocks. On the other hand, since from these measurements only dose rates and effective dose rates outdoors can be determined, additional studies are also required in order to be able to determine the dose rates and effective dose rates indoors. This would provide the possibility to accurately determine the public total effective dose rates due to natural radioactivity.

## Acknowledgements

This work is conducted with financial support from the Cyprus Research Promotion Foundation (Grant No. 45/2001). We would also like to thank G. Petrides and I. Panagides from the Cyprus Geological Survey Department of the Ministry of Agriculture, Natural Sources and Environment in Nicosia, for assisting us with their expertise in geological issues.

## TABLE CAPTIONS

**Table 1.** Samples categorization and rock nomenclature as well as natural radioactivity concentrations of $^{232}$Th, $^{238}$U and $^{40}$K in the measured samples.

| Rock Type | | Concentration ± Stat. Error (*Bq kg*$^{-1}$) | | |
|---|---|---|---|---|
| | | **$^{232}$Th** | **$^{238}$U** | **$^{40}$K** |
| **OPHIOLITIC ORIGIN** | | | | |
| **1** | Mathiatis Pyrite Mine Tippings | 2.5 ± 1.1 | 50.5 ± 0.6 | 31.9 ± 16.8 |
| **2** | Troodos Upper Pillow Lavas (Olivine Basalt, Picrite Basalt) | 3.2 ± 0.2 | 2.8 ± 0.1 | 894.2 ± 34.0 |
| **3** | Troodos Lower Pillow Lavas (Oversaturated Basalt) | 2.9 ± 0.2 | 5.2 ± 0.3 | 307.2 ± 11.8 |
| **4** | Sheeted Dyke Complex (Diabase) | 2.7 ± 0.1 | 2.7 ± 0.1 | 106.6 ± 4.2 |
| **5** | Gabbro | 1.8 ± 0.1 | 1.5 ± 0.6 | 71.2 ± 2.9 |
| **6** | Plagiogranite | 2.8 ± 0.1 | 3.0 ± 0.1 | 128.4 ± 5.0 |
| **7** | Asbestos Mine Waste Tip Material | 1.5 ± 0.1 | 1.7 ± 0.1 | 20.0 ± 0.9 |
| **8** | Harzburgite | 1.7 ± 0.1 | 1.2 ± 0.1 | 16.2 ± 0.8 |
| **9** | Wehrlite | 1.4 ± 0.1 | 0.9 ± 0.1 | 13.0 ± 0.6 |
| **10** | Dunite | 1.3 ± 0.1 | 1.3 ± 0.1 | 16.9 ± 0.9 |
| **11** | Pyroxenite | 1.8 ± 0.1 | 1.1 ± 0.1 | 14.9 ± 0.7 |
| | | **Mean ± S.D.** | **Mean ± S.D.** | **Mean ± S.D.** |
| | | **2.1 ± 0.2** | **6.5 ± 4.4** | **147.3 ± 79.3** |
| **SEDIMENTARY ORIGIN** | | | | |
| **12** | Limestone | 2.1 ± 0.1 | 8.3 ± 0.3 | 20.0 ± 1.0 |
| **13** | Red Clay Soil (Terra Rossa) | 38.3 ± 0.5 | 21.0 ± 0.5 | 438.5 ± 10.1 |
| **14** | Pediaios River Alluvium | 13.4 ± 0.2 | 17.6 ± 1.4 | 261.3 ± 1.5 |
| **15** | Larnaka Beach Deposits | 14.9 ± 0.5 | 13.2 ± 0.5 | 127.7 ± 5.2 |
| **16** | Lemesos Beach Deposits | 3.9 ± 0.2 | 3.1 ± 0.1 | 76.1 ± 3.0 |
| **17** | Klavdhia, Raised Marine Terrace Deposits | 21.3 ± 0.7 | 90.3 ± 2.5 | 240.3 ± 9.5 |
| **18** | Celestite | 2.4 ± 0.2 | 56.0 ± 1.5 | 17.6 ± 1.0 |
| **19** | Gypsum | 2.8 ± 0.2 | 3.8 ± 0.1 | 20.9 ± 1.1 |
| **20** | Chalk | 9.6 ± 0.3 | 5.4 ± 0.2 | 87.0 ± 3.5 |
| **21** | Melange | 52.8 ± 1.5 | 16.0 ± 0.5 | 392.5 ± 15.0 |
| **22** | Stratified Mamonia | 20.0 ± 0.6 | 7.9 ± 0.3 | 245.3 ± 9.5 |
| **23** | Montmorilonite (Bentonitic Clay) | 40.7 ± 1.1 | 18.3 ± 0.6 | 278.9 ± 10.8 |
| **24** | Marl (Lefkosia center) | 7.1 ± 0.3 | 8.9 ± 0.3 | 187.3 ± 7.3 |
| **25** | Calcareous Sandstone | 8.7 ± 0.3 | 21.7 ± 0.6 | 53.5 ± 2.4 |
| **26** | Marl (Dhali area) | 10.5 ± 0.4 | 19.1 ± 0.6 | 329.6 ± 12.8 |
| **27** | Marl (Lefkosia suburbs) | 30.8 ± 0.9 | 41.2 ± 1.2 | 433.6 ± 16.7 |
| **28** | Paphos Beach Deposits | 4.8 ± 0.2 | 6.3 ± 1.2 | 147.3 ± 1.6 |
| | | **Mean ± S.D.** | **Mean ± S.D.** | **Mean ± S.D.** |
| | | **16.7 ± 3.7** | **21.0 ± 5.5** | **197.5 ± 35.2** |

**Table 2.** Conversion factors for different radionuclides as deduced by Beck et al., Saito and Jacob, in comparison with recent results from various Monte Carlo techniques, obtained by Clouvas et al. The calculated values are given in units of *nGy* $h^{-1}$ $Bq^{-1}$ *kg*.

| **Nuclide** | **BECK** | **SAITO** | **MCNP** | **MC** | **GEANT** |
|---|---|---|---|---|---|
| **$^{232}$Th series** | | | | | |
| $^{228}$Ac | 0.27800 | 0.22100 | 0.18526 | 0.19120 | 0.19594 |
| $^{212}$Bi | 0.02120 | 0.02720 | 0.02256 | 0.02305 | 0.02383 |
| $^{224}$Ra | - | 0.00214 | 0.00156 | 0.00167 | 0.00167 |
| $^{208}$Tl | 0.32100 | 0.32600 | 0.28944 | 0.28871 | 0.30312 |
| $^{212}$Pb | 0.02120 | 0.02770 | 0.01796 | 0.01926 | 0.01917 |
| **Total** | 0.66600 | 0.60400 | 0.51678 | 0.52389 | 0.54373 |
| **$^{238}$U series** | | | | | |
| $^{214}$Pb | 0.04720 | 0.05460 | 0.04150 | 0.04413 | 0.04342 |
| $^{226}$Ra | - | 0.00125 | 0.00092 | 0.00099 | 0.00100 |
| $^{214}$Bi | 0.37800 | 0.40100 | 0.33849 | 0.34156 | 0.35554 |
| **Total** | 0.43000 | 0.46300 | 0.38092 | 0.38668 | 0.39996 |
| **$^{40}$K** | 0.04220 | 0.04170 | 0.03780 | 0.03808 | 0.03995 |

**Table 3.** Average activity, dose and effective dose rates assessment for the $^{232}$Th and $^{238}$U series and $^{40}$K, as obtained from the analysis of the spectrum shown in Fig. 2.

| | **DRCF** *(nGy h$^{-1}$ per Bq kg$^{-1}$)* | **Nuclide** | **Average activity ± Stat. Err.** *(Bq kg$^{-1}$)* | **Dose rates outdoors** *(nGy h$^{-1}$)* | **Average dose rates outdoors** *(nGy h$^{-1}$)* | **Effective dose rates outdoors** *(μSv y$^{-1}$)* |
|---|---|---|---|---|---|---|
| **Thorium Series** | 0.52813 | $^{228}$Ac (911 keV) | 32.3 ± 1.4 | 17.0 | 13.9 | 17.1 |
| | | $^{212}$Bi (727 keV) | 20.4 ± 1.6 | 10.8 | | |
| | | $^{208}$Tl (583 keV) | 8.2 ± 0.4 | 4.3 | | |
| | | $^{212}$Pb (238 keV) | 44.4 ± 1.8 | 23.4 | | |
| **Uranium Series** | 0.38919 | $^{214}$Pb (352 keV) | 18.2 ± 0.8 | 7.1 | 6.6 | 8.1 |
| | | $^{214}$Bi (609 keV) | 15.7 ± 0.7 | 6.1 | | |
| **Potassium** | 0.03861 | $^{40}$K (1460 keV) | 257.0 ± 10.8 | 9.9 | 9.9 | 12.1 |
| | | | | **Total** | **30.4** | **37.3** |

**FIGURE CAPTIONS**

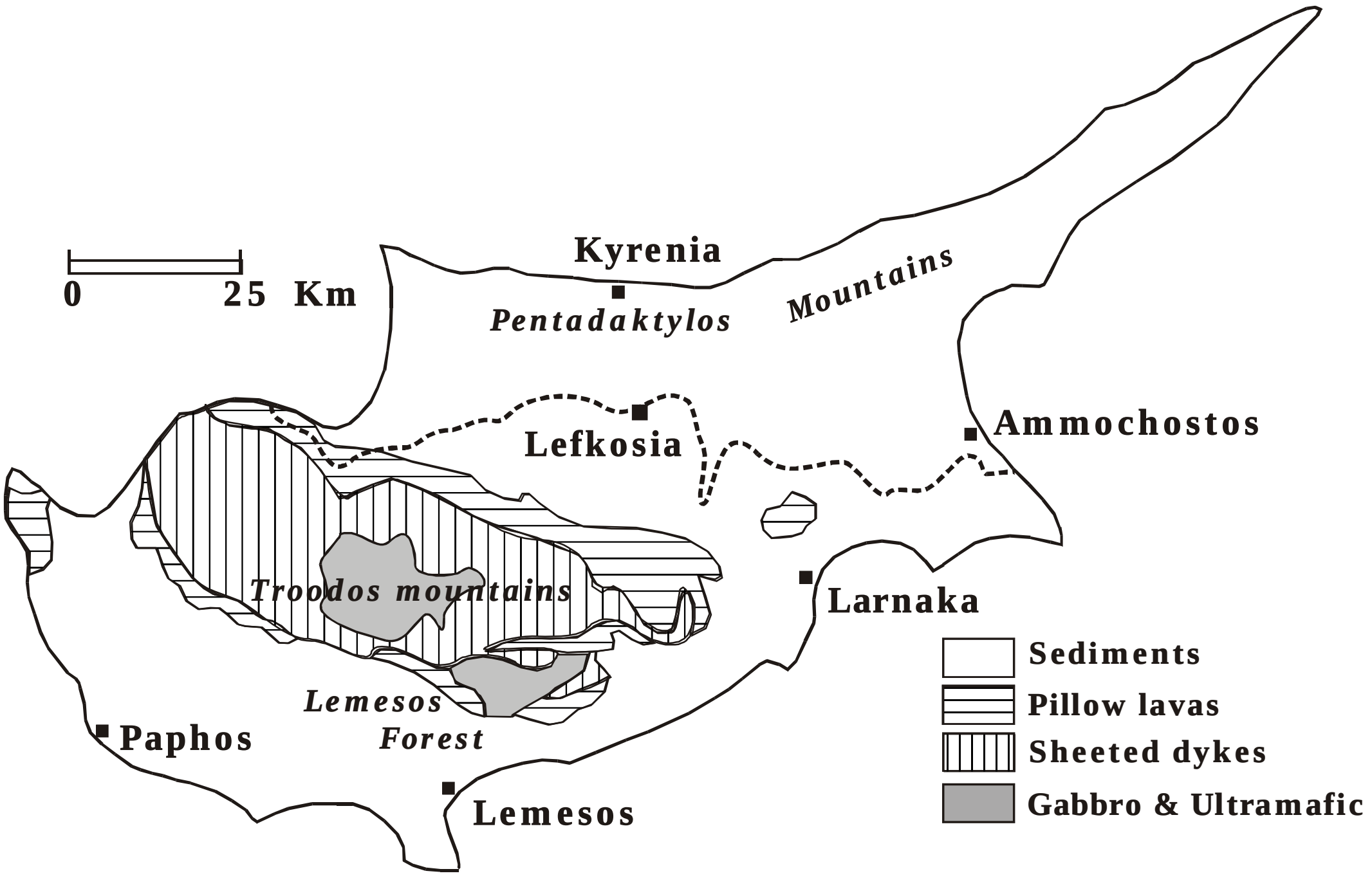

**Figure 1**. General geological map of Cyprus with emphasis on the Troodos ophiolitic complex, which mainly consists of four layers: sediments, pillow lavas, sheeted dykes and gabbro, diabase and ultramafics. Geological formations located in the northern part of Cyprus, such as those consisting the Pentadaktylos mountains, are inaccessible because of the division (marked by the dotted line) of the island by the Turkish occupation forces since 1974.

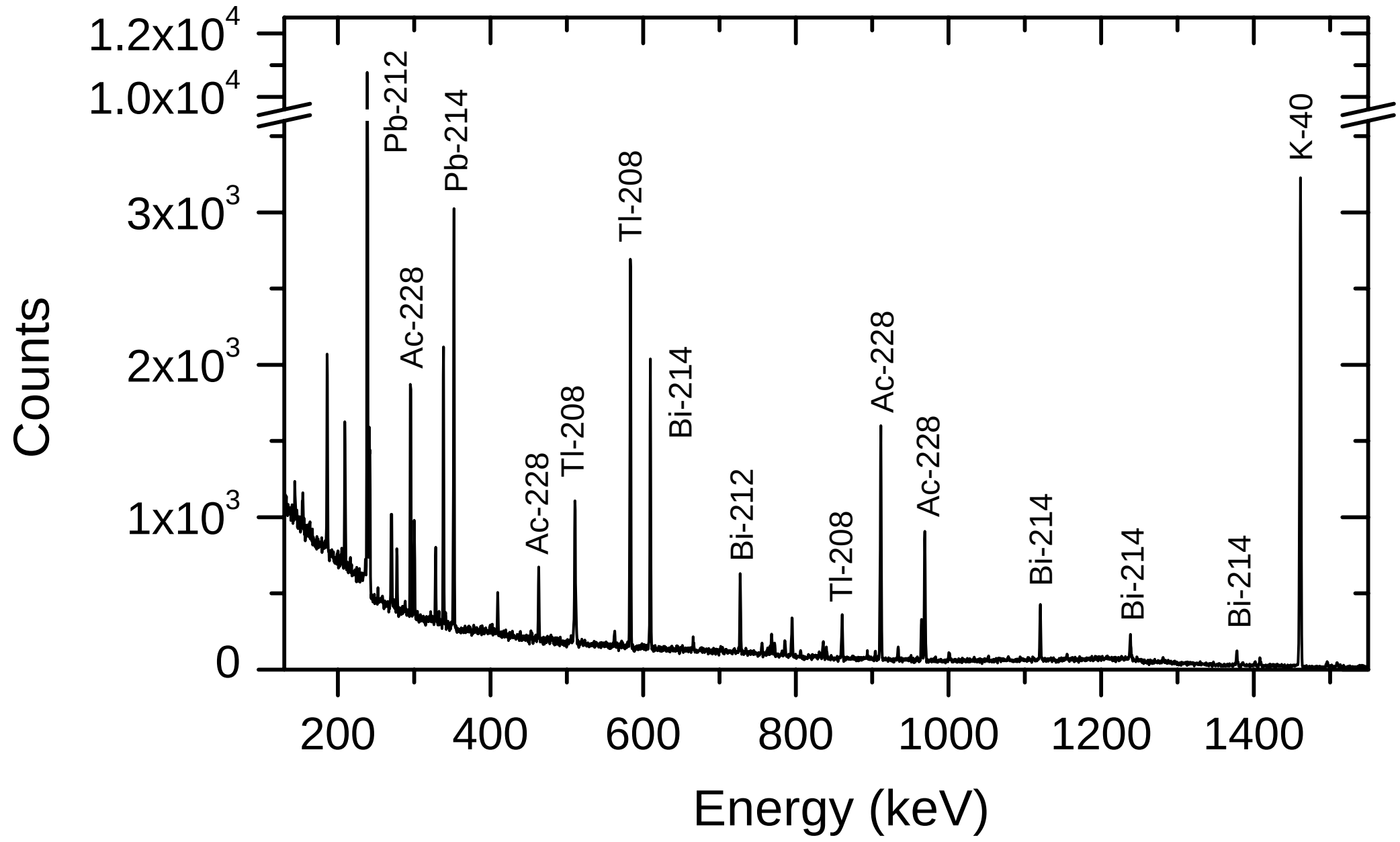

**Figure 2**. The main part of a typical γ–ray spectrum measured for sample 23 (Montmorilonite – Bentonitic clay). The important identified photopeaks and their associated radionuclides are shown.

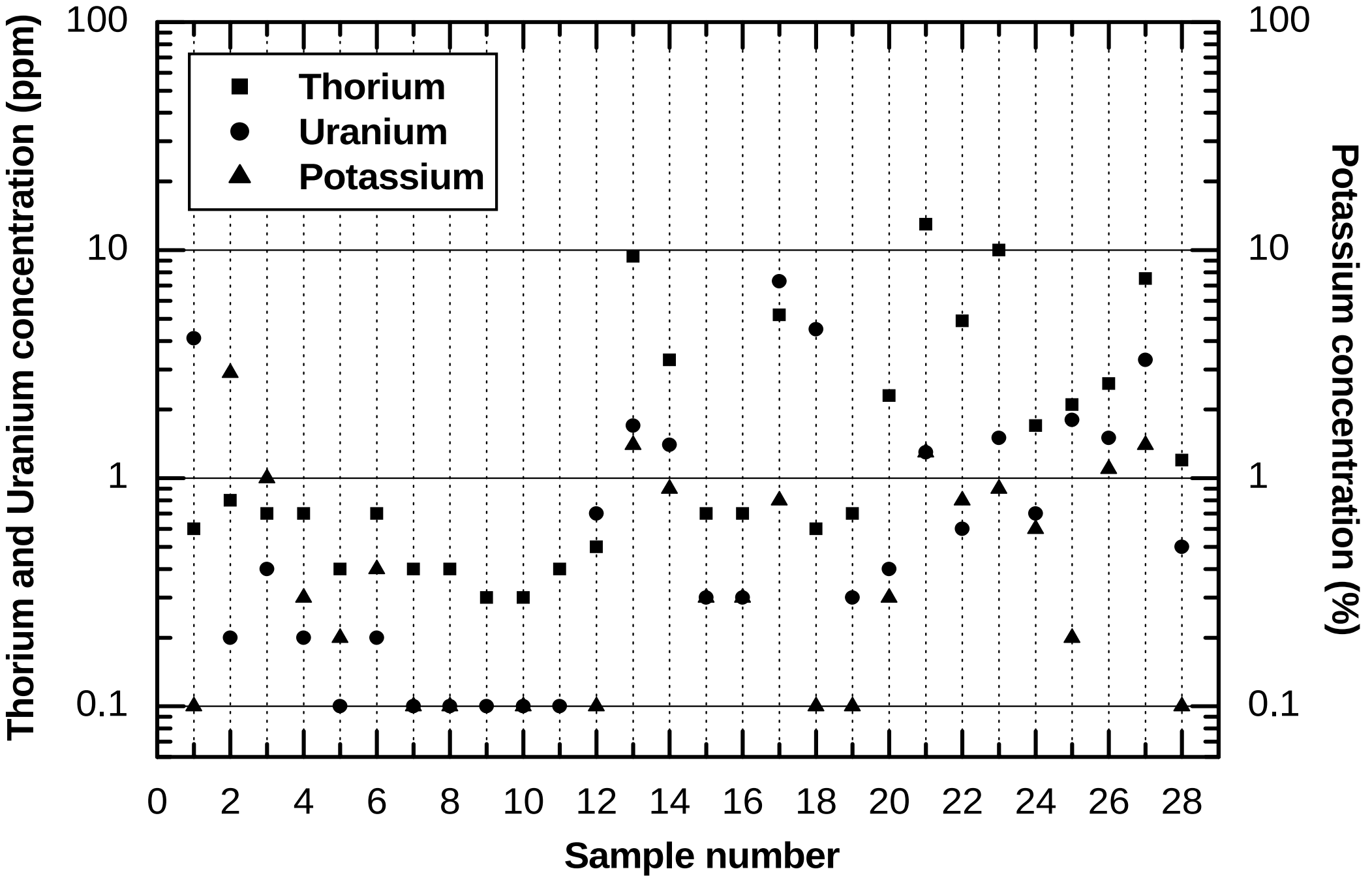

**Figure 3.** Thorium, uranium and potassium elemental concentrations in samples of the characteristic rock types present in Cyprus. The sample numbers correspond to the sample description given in table 1.

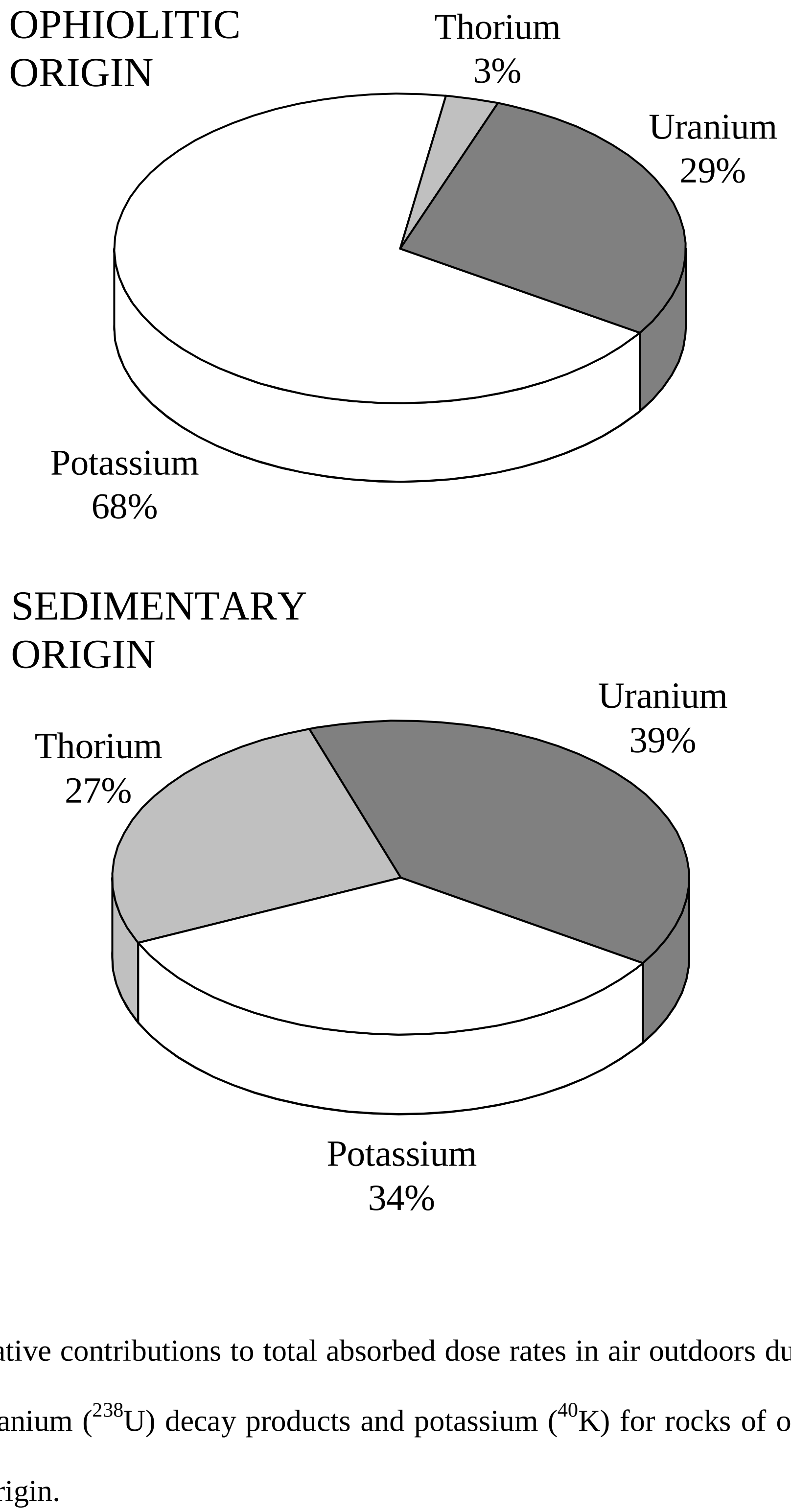

**Figure 4.** Relative contributions to total absorbed dose rates in air outdoors due to thorium ($^{232}Th$) and uranium ($^{238}U$) decay products and potassium ($^{40}K$) for rocks of ophiolitic and sedimentary origin.